# Simulating the Charge Dispersion Phenomena in Micro Pattern Gas Detectors with a Resistive Anode

M. S. Dixit[*a, b] and A. Rankin[a]

[a]Department of Physics, Carleton University
1125 Colonel By Drive, Ottawa, ON, K1S 5B6, Canada
and
[b]TRIUMF, Vancouver, BC, Canada

**Abstract:**

The TPC for the International Linear Collider (ILC) will need to measure about 200 track points with a transverse resolution close to 100 μm. The resolution goal is beyond the capability of the conventional proportional wire/cathode pad TPC and Micro-Pattern Gas Detectors (MPGD) are being developed to meet the challenge. The standard MPGD readout techniques will, however, have difficulty in achieving the ILC-TPC resolution goal with the 2 x 6 mm$^2$ wide pads as was initially envisioned. Proposals for smaller width pads will improve the resolution but will require a larger number of readout channels and increase the TPC detector cost and complexity. The new MPGD readout concept of charge dispersion has the potential to achieve the ILC-TPC resolution goal without resorting to narrower pads. This was recently demonstrated in cosmic ray tests of a small prototype TPC read out with MPGDs using the charge dispersion technique. Here we describe the simulation of the charge dispersion phenomena for the MPGD-TPC. The detailed simulation includes initial ionization clustering, electron drift, diffusion effects, the intrinsic detector pulse-shape and electronics effects. The simulation is in excellent agreement with the experimental data and can be used to optimize the MPGD charge dispersion readout for the TPC.

## 1. Introduction

The Time Projection Chamber (TPC) [1] is a prime candidate for the main charged particle tracker at the future International Linear Collider (ILC). The ILC-TPC operating in a ~ 4 T magnetic field should measure 200 track points with a transverse resolution 100 μm for drift lengths in excess of 2 m. The resolution goal is near the fundamental limit from the ionization electron statistics and transverse diffusion in the TPC gas, and beyond the capability of the conventional proportional wire/cathode pad TPC [2].

A TPC read out with Micro Pattern Gas Detectors (MPGD), such as the Gas Electron Multiplier (GEM) [3] and the Micromegas [4], has the potential to reach the ILC resolution goal. In normal tracking applications, MPGDs typically achieve 40 - 50 μm resolution [5] with ~ 200 μm pitch

---

anode pads. In conventional proportional wire TPCs, much wider cathode pads are read out to compute the charge centroid, which determines the avalanche r-φ coordinate with precision. The fundamental resolution limit for the wire TPC comes from the **E×B** and track angle systematic effects [6,7]. These effects are negligible for the MPGD readout as the scale over which the **E** and **B** fields are not parallel is much smaller and also there is no preferred direction in contrast to the wire readout. MPGD anode pads, 2 x 6 mm$^2$ in size were initially proposed for the ILC-TPC readout [8]. With the strong suppression of transverse diffusion in high magnetic fields, the track ionization charge clusters arriving at the TPC endplate will then often be confined to only one or two pads. This makes the anode pad centroid determination difficult and results in loss of resolution. Reducing the pad width to improve resolution will require a larger number of readout channels and increase the detector cost and complexity.

In an attempt to improve the charge centroid determination and hence the spatial resolution for wide pads, a novel concept of position sensing from charge dispersion [9] has been developed where the MPGD anode is made of a thin high surface resistivity film. The resistive anode is bonded to the readout plane with an insulating layer of glue, which acts as a dielectric spacer between the two planes. The composite anode-readout pad plane structure forms a distributed 2-dimensional resistive-capacitive network. Any localized charge arriving at the anode surface will be dispersed with the RC time constant determined by the anode surface resistivity and the capacitance per unit area, the latter determined by the spacing between the anode and readout planes and the dielectric constant of the glue. With the avalanche charge dispersing and covering a larger number of pads with time, wider pads can be used for position determination. The charge dispersion process can be completely described by material properties and geometry and, in contrast to diffusion which is statistical in nature, there is no loss of accuracy in determining the centroid of a wider distribution. Fig. 1 shows the schematics of the double GEM test cell used in our initial tests of the charge dispersion readout concept.

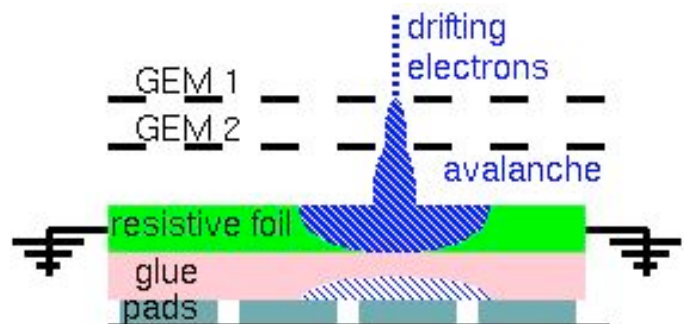

Fig. 1. Schematic of a double GEM test cell designed for charge dispersion studies.

The first proof of principle tests of charge dispersion for the GEM were carried out using a collimated soft x-ray source and have been previously published [9]. These were followed by cosmic ray resolution studies of a prototype TPC read out with GEM [10] and with Micromegas [11] using the charge dispersion technique. In this paper, we present the results of a detailed simulation of the charge dispersion phenomenon based on the model described in [9]. The charge dispersion effect is first calculated for a single point-like charge cluster deposited instantaneously on the resistive anode. The finite extent of the charge cluster due to longitudinal and transverse diffusion, the effects of intrinsic rise-time of the MPGD charge pulse and the rise- and fall-time effects in electronics are then included. Track signals can be generated by summing signals due to individual charge clusters along the track. The simulation is in excellent agreement with the observed features of charge dispersion and can be used to optimize the charge dispersion readout system parameters for TPC.

## 2. Modeling the charge dispersion phenomena

If a charge is deposited on the resistive anode, the equation describing the time evolution of the surface charge density function on the two dimensional continuous RC network is given by [9]:

$$\frac{\partial \rho}{\partial t} = h\left(\frac{\partial^2 \rho}{\partial x^2} + \frac{\partial^2 \rho}{\partial y^2}\right), \tag{1}$$

where $h = 1/RC$.

The solution to equation 1 for a resistive anode of finite size is an infinite Fourier series. A closed form solution becomes possible, however, for the case of a delta function point charge deposited at $x = y = t = 0$ and when the edges are at infinity:

$$\rho_\delta(x,y,t) = \left(\frac{1}{2\sqrt{\pi t h}}\right)^2 \exp\left[-(x^2 + y^2)/4th\right]. \tag{2}$$

The true initial charge profile is not a delta function but has a finite size and can be described by a Gaussian with a width determined by transverse diffusion. For a cluster with charge $Nq_e$, the anode surface charge density as a function of space and time is obtained by convoluting equation 2 with the Gaussian describing the finite charge cluster of width $w$:

$$\rho(x,y,t) = \frac{Nq_e}{2\pi(2ht + w^2)} \exp\left[-(x^2 + y^2)/(2(2ht + w^2))\right]. \tag{3}$$

The charge on a pad can be found by integrating the charge density function over the pad area:

$$Q_{pad}(t) = \frac{Nq_e}{4}\left[erf(\frac{x_{high}}{\sqrt{2}\sigma_{xy}}) - erf(\frac{x_{low}}{\sqrt{2}\sigma_{xy}})\right]\left[erf(\frac{y_{high}}{\sqrt{2}\sigma_{xy}}) - erf(\frac{y_{low}}{\sqrt{2}\sigma_{xy}})\right]. \tag{4}$$

where $x_{low}, x_{high}, y_{low}, y_{high}$ define the pad boundaries, and $\sigma_{xy} = \sqrt{2th + w^2}$.

The charge is also not deposited instantaneously. The detector pulse has a finite intrinsic rise time and the signal is also affected by electron arrival time spread due to longitudinal diffusion. To compare to experiment, the characteristics of the front-end charge preamplifiers need also to be included. The parameterization of these time dependent effects is described below.

*The intrinsic rise-time of the detector charge pulse*: From Ramo's theorem [12], the charge pulse on the GEM anode rises linearly with time as the electrons drift in the induction gap toward the anode at a constant drift velocity.

$$\left.\begin{array}{ll} R(t) = t/T_{rise} & \text{for } 0 < t < T_{rise} \\ \phantom{R(t)} = 1 & \text{for } t > T_{rise} \\ \phantom{R(t)} = 0 & \text{for } t < 0 \end{array}\right\} \tag{5}$$

where $T_{rise}$ is the time, taken by an electron cluster to traverse the GEM induction gap.

*Longitudinal diffusion:* Longitudinal diffusion increases the size of electron charge clusters in the drift direction. The arrival time distribution of electrons at the resistive anode is a Gaussian with standard deviation $\sigma$ determined by the longitudinal diffusion coefficient for the gas and the drift time. The longitudinal cluster profile takes the form:

$$L(t) = \frac{1}{\sigma\sqrt{2\pi}} \exp(-t^2/2\sigma^2). \tag{6}$$

*Electronics shaping time effects:* The timing characteristics of the front-end charge preamplifier are included into the charge dispersion model as follows. The step response of a charge amplifier consists of an exponential rise, $t_r$, and an exponential decay, $t_f$ described by equation 7.

$$\left. \begin{array}{ll} A(t) = \exp(-t/t_f)[1-\exp(t/t_r)] & \text{for } t > 0 \\ \\ = 0 & \text{for } t < 0 \end{array} \right\} \tag{7}$$

The individual time effects are combined to create a single function describing the time dependence of the model. The convolution of $R(t)$, $L(t)$ and $A(t)$ yields:

$$I(t) = \frac{1}{2T_{rise}} \begin{bmatrix} \exp(\sigma^2 a^2/2 - at)\left[\mathrm{erf}\left(\frac{t - T_{rise} - \sigma^2 a}{\sigma\sqrt{2}}\right) + 1\right] - \\ \exp(\sigma^2 b^2/2 - bt)\left[\mathrm{erf}\left(\frac{t - T_{rise} - \sigma^2 b}{\sigma\sqrt{2}}\right) + 1\right] + \\ \exp(\sigma^2 a^2/2 - a(t - T_{rise}))\left[\mathrm{erf}\left(\frac{t - 2T_{rise} - \sigma^2 a}{\sigma\sqrt{2}}\right) + 1\right] - \\ \exp(\sigma^2 b^2/2 - b(t - T_{rise}))\left[\mathrm{erf}\left(\frac{t - 2T_{rise} - \sigma^2 b}{\sigma\sqrt{2}}\right) + 1\right] + \\ \exp(\sigma^2 a^2/2 - at)\left[\mathrm{erf}\left(\frac{t - \sigma^2 a}{\sigma\sqrt{2}}\right) + \mathrm{erf}\left(\frac{t - T_{rise} - \sigma^2 a}{\sigma\sqrt{2}}\right)\right] - \\ \exp(\sigma^2 b^2/2 - bt)\left[\mathrm{erf}\left(\frac{t - \sigma^2 b}{\sigma\sqrt{2}}\right) + \mathrm{erf}\left(\frac{t - T_{rise} - \sigma^2 b}{\sigma\sqrt{2}}\right)\right] \end{bmatrix}. \tag{8}$$

where $a = 1/t_f$; $b = 1/t_f + 1/t_r$.

The convolution of $I(t)$ and $Q_{pad}(t)$ results in the full theoretical model, which is compared to the experiment. This convolution is handled numerically.

## 3. Simulation – Results

At the basic level, the model describes the charge dispersion phenomena for a single charge cluster. The model was first tested by comparing simulation to the measurements made using collimated soft x rays from a copper target x-ray tube [9]. X-ray photon conversions in the gas produce initial ionization charge clusters. A double-GEM test cell was modified for the charge dispersion studies. A 25 μm thick Carbon-loaded Kapton foil with a surface resistivity of 530 kΩ per square was bonded to the readout pad plane with 50 μm thick insulating layer of glue. The capacitance density of the anode with respect to the readout plane was about 0.21 pF/mm$^2$ with the glue acting as a dielectric spacer. The measurements were carried out with the test cell filled with Ar/$CO_2$ 90/10 gas. The gas gain was about 3500. The readout plane, shown in Fig 5 (a), had five rows of twelve pads, 2 x 6 mm$^2$ each. The front-end electronics and the data acquisition system were the same as described in [9].

With the x-ray tube run at 6.5 kV, the average photon energy was about 4.5 keV as low energy bremsstrahlung photons from the x-ray tube were absorbed by the x-ray tube and detector windows. A ~ 40 μm diameter hole in a thin brass disk in front of the detector produced a reduced pinhole image of the x-ray tube focal spot in the GEM drift gap. The GEM test cell geometry, the electric fields, and the diffusion properties of the gas determine the physical dimensions of the electron avalanche charge cluster arriving at the resistive anode. For the x-ray measurements reported here, we estimate that the charge cluster arriving at the anode had a transverse RMS width of about 250 μm and a longitudinal RMS spread in time of about 13 ns.

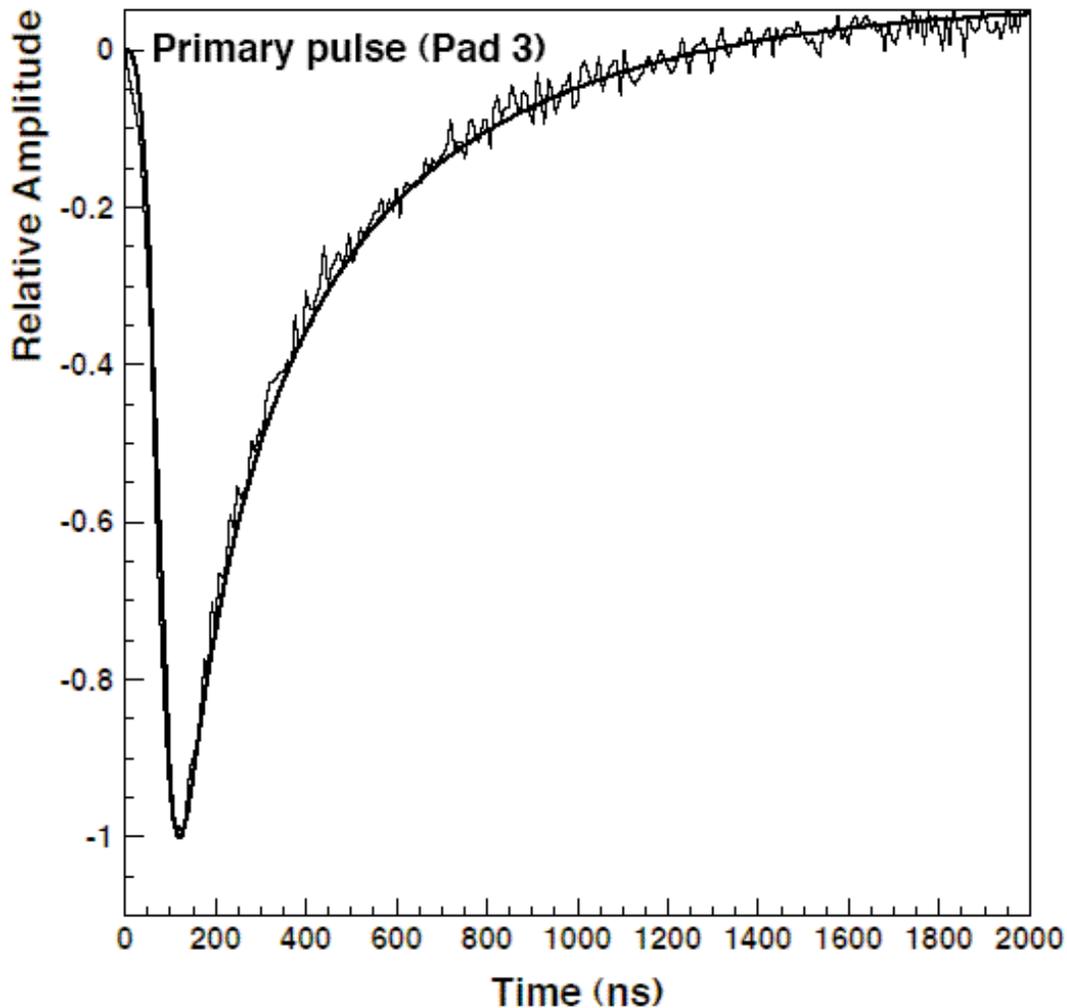

Fig. 2. Simulated and observed primary pulse for the x-ray spot centered on pad 3 – see Fig. 5 for the pad layout. The simulation is normalized to the experiment.

The simulation was used to compute the detailed shapes and amplitudes of charge dispersion pulses. As an example, Fig. 2 shows the measured and the simulated primary charge pulse for pad 3 with the x-ray spot centered on the pad – see Fig 5(a) for the pad layout. The amplitude of the simulated pulse is normalized to the measured primary pulse. With the same normalization, the simulation reproduces the detailed shape and the amplitude of the observed secondary pulse on the next neighbor pad shown in Fig. 3. The small difference between the measured and the simulated pulse is due to the existence of induced pulse. We have previously reported on the measurement of induced pulses in GEM detectors [13,14]. The induced pulses were not included in the simulation as they deposit zero net charge on the pads and do not contribute to the charge dispersion signal.

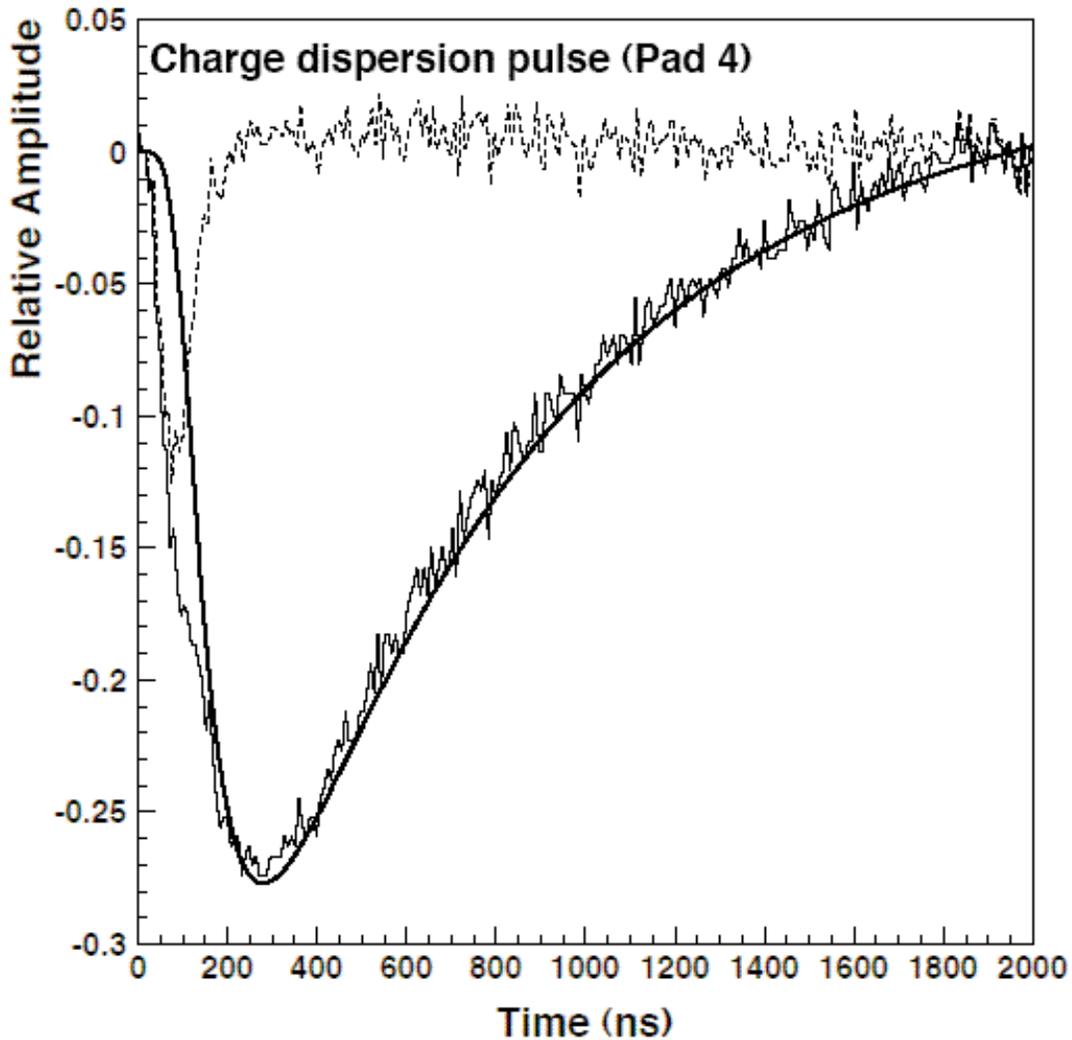

Fig. 3. Measured and simulated secondary pulses on Pad 4 from charge dispersion with the x-ray spot centered on Pad 3. The normalization determined for the primary pulse on Pad 3 was used to normalize the simulated secondary pulse. The difference (dotted line) between the simulation and the experiment (solid line) comes from the induced signal due to electron drift in the GEM induction gap. Induced signals were not simulated as they deposit zero net charge and do not affect charge dispersion signals.

The simulation was next used to generate single cluster pad response function (PRF) for charge dispersion. Pad signals were first generated by incrementing the charge cluster position across the pad width in small steps. The theoretical PRF is determined from simulated pulse amplitudes as a function of cluster position with respect to the pad centre. Fig. 4 shows the theoretical and the measured PRFs for pads 2, 3 and 4 for the pad layout shown in Fig. 5 (a). The theoretical PRF is in good agreement with the data. The small differences can be attributed to point-to-point variations of capacitance per unit area and the anode surface resistivity.

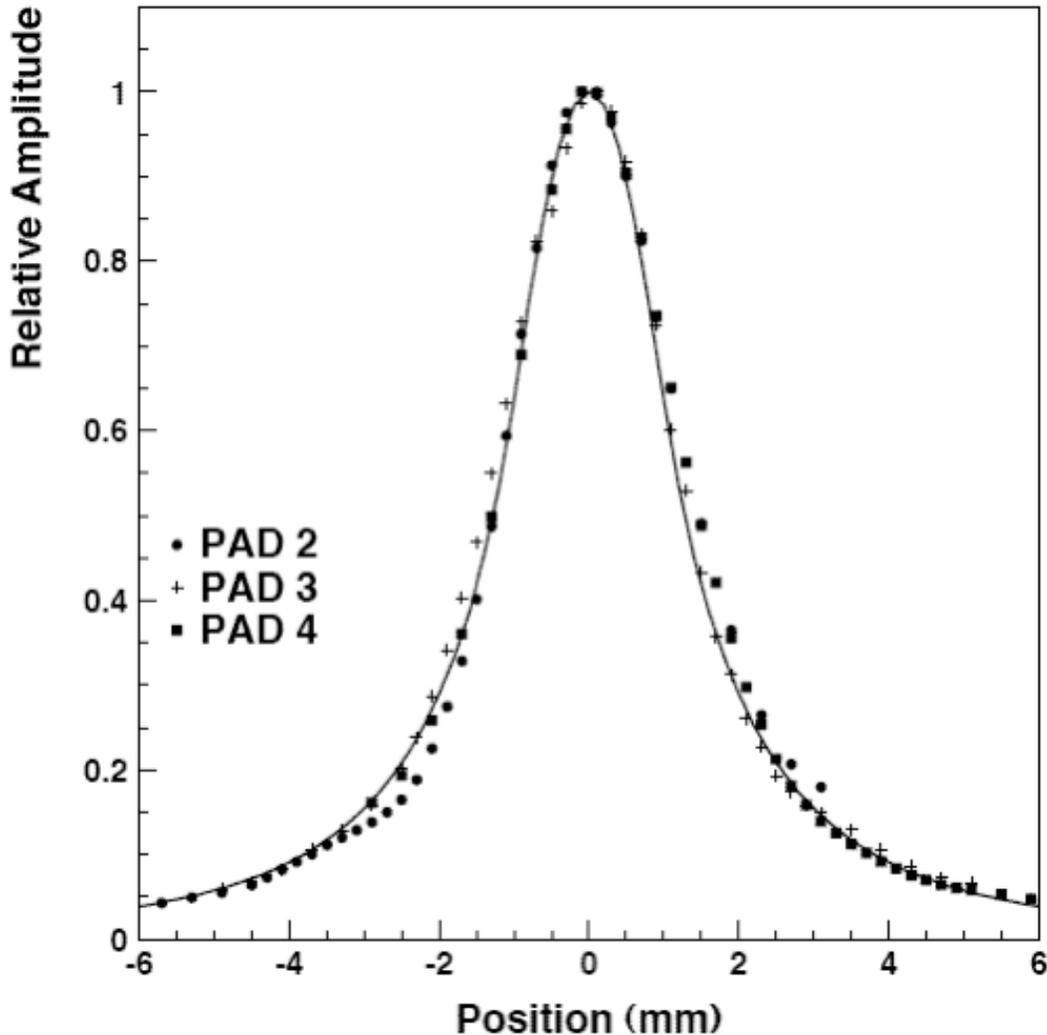

Fig. 4 Simulated pad response functions (PRF) compared to the measured PRFs for pads 2, 3 and 4. The pad layout is shown in Fig. 5 (a). The differences between the simulation and the measured PRFs are attributed to local variations in the system RC constant. Nevertheless, the simulated PRF is in good agreement with the data.

The model of charge dispersion was also tested for particle tracks using cosmic ray data [10] for a small GEM readout TPC with 15 cm long drift region. A sample track is shown in Fig. 5. Pad signals for a particle track are simulated by summing the response of the pad to the ionization charge clusters produced by the track. Since we have no detailed information on the cluster size distribution for a particular track, the clustering effect was taken into account in an approximate way. Equal size equally spaced charge clusters were generated along the fitted track with the sum of cluster charges over each pad set equal to the measured pad charge. The transverse and longitudinal dimensions of the clusters at the readout were determined from diffusion computed for the track geometry for the event. Simulated pulses were generated for pads in the middle row and compared to the measured pulses. The simulation is in excellent agreement with the experiment as can be seen in Fig. 5.

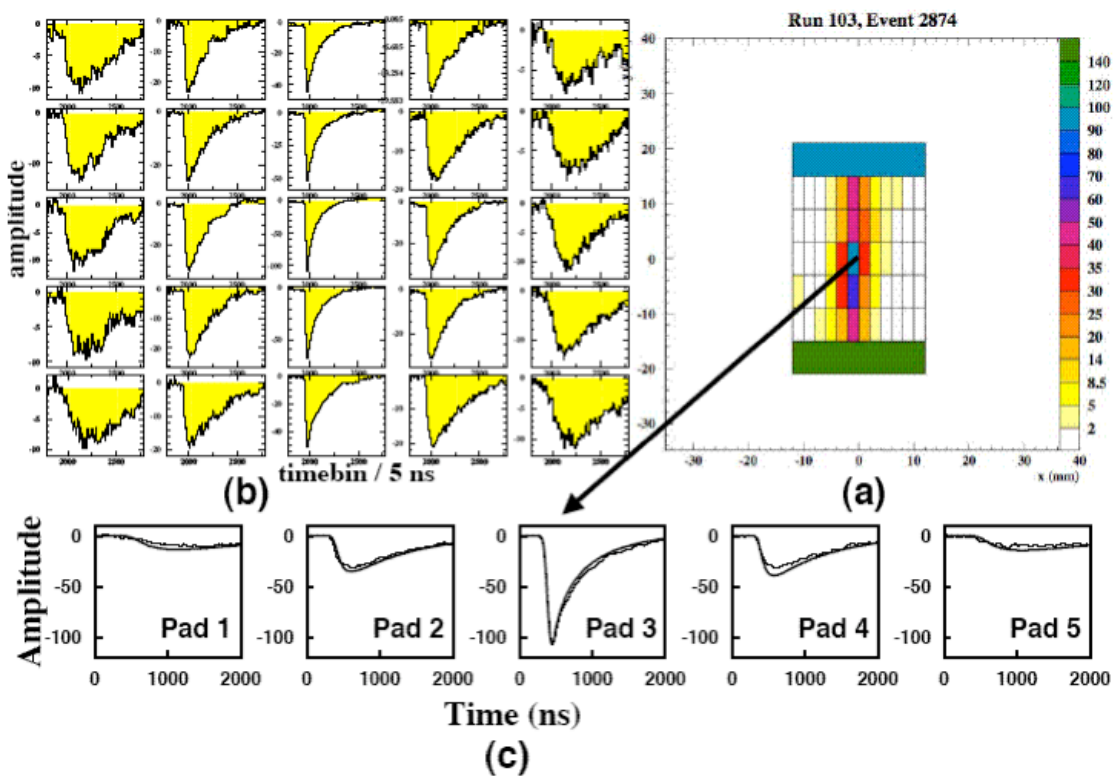

Fig. 5. (a) The schematics of the pad layout used for the x-ray and TPC cosmic ray tests. The central 5 row of pads with 2 x 6 mm$^2$ pads were used for tracking and the two outer rows with single long pads were used for triggering. (b) Cosmic ray track signals on the five middle pads in each of the tracking rows. (c) Measured and simulated cosmic ray track signals for the five middle pads in the central tracking row.

## 4. Conclusions

A detailed physical model has been developed to describe the charge dispersion phenomena in MPGDs with a resistive anode. The simulation developed for this study provides an excellent description of the experimental measurements and can therefore be used as a basis for a full simulation of a TPC with charge dispersion readout. Although the charge dispersion simulation and experimental measurements reported in this paper are for the GEM TPC readout, the model is equally applicable to the Micromegas.

## 5. Acknowledgements:

We thank Alain Bellerive for carefully reading the manuscript and for helpful suggestions. This research was supported by a project grant from the Natural Sciences and Engineering Research Council of Canada. TRIUMF receives federal funding via a contribution agreement through the National Research Council of Canada.

**References:**

1. D. R. Nygren, A Time Projection Chamber - 1975, Presented at 1975 PEP Summer Study, PEP 198 (1975), and Included in Proceedings.
2. W. Blum and L. Rolandi, Particle detection with Drift Chambers, Springer Berlin, 1993, pp. 179—184.
3. F. Sauli, Nucl. Instrum. Meth., A386 (1997) 531.
4. Y. Giomataris et al., Nucl. Instrum. Meth., A376 (1996) 29.
5. F. Sauli and A. Sharma, Ann. Rev. Nucl. Part. Sci. 49 (1999) 341.
6. C. K. Hargrove et al., Nucl. Instrum. Meth., A219 (1984) 461.
7. S. R. Amendolia et al., Nucl. Instrum. Meth., A283 (1989) 573.
8. Worldwide Study of the Physics and Detectors for Future $e^+e^-$ Linear Colliders (http://physics.uoregon.edu/~lc/randd.html).
9. M. S. Dixit et al., Nucl. Instrum. Meth., A518 (2004) 689.
10. R.K.Carnegie et al,'"First Tracking Experience for MPGD TPC readout with Charge Dispersion on a Resistive Anode, Proceedings of International Conference on Linear $e^+e^-$ Colliders, LCWS2004, Paris (2004).
11. A. Bellerive et al., Spatial Resolution of a Micromegas TPC Using the Charge Dispersion Signal, Proceedings of International Linear Collider Workshop, LCWS2005, Stanford, USA, arXiv:physics/0510085.
12. S. Ramo, Proc. IRE, 27 (1939) 584.
13. M.S. Dixit et al, Proceedings of Workshop on Micro-Pattern Gas Detectors, Orsay France (1999).
14. D. Karlen et al, Physics and experiments with future linear $e^+e^-$ colliders, LCWS2000, American Institute of Physics Conf. Proc. Vol 578.